\documentclass[aps,pra,twocolumn,notitlepage,superscriptaddress]{revtex4-2}

\usepackage{physics}
\usepackage{xcolor,graphicx}
\usepackage{comment}
\usepackage{siunitx}

\usepackage{amsthm,amssymb}
\usepackage{array}
\usepackage{hyperref}
\usepackage{mathtools}
\usepackage{bm,times,enumitem}
\usepackage{physics}
\usepackage{algorithm}
\usepackage{algc}
\usepackage{algcompatible}
\usepackage{algpseudocode}
\usepackage{soul}
\usepackage{dsfont}

\begin{document}

\title{Benchmarking Bayesian quantum estimation}

\author{Valeria Cimini}
\email{valeria.cimini@uniroma1.it}
\affiliation{Dipartimento di Fisica, Sapienza Universit\`{a} di Roma, Piazzale Aldo Moro 5, I-00185 Roma, Italy}

\author{Emanuele Polino}
\affiliation{Dipartimento di Fisica, Sapienza Universit\`{a} di Roma, Piazzale Aldo Moro 5, I-00185 Roma, Italy}
\affiliation{Centre for Quantum Dynamics and Centre for Quantum Computation and Communication Technology, Griffith University, Yuggera Country, Brisbane, Queensland 4111, Australia}

\author{Mauro Valeri}
\affiliation{Dipartimento di Fisica, Sapienza Universit\`{a} di Roma, Piazzale Aldo Moro 5, I-00185 Roma, Italy}

\author{Nicol\`o Spagnolo}
\affiliation{Dipartimento di Fisica, Sapienza Universit\`{a} di Roma, Piazzale Aldo Moro 5, I-00185 Roma, Italy}

\author{Fabio Sciarrino}
\affiliation{Dipartimento di Fisica, Sapienza Universit\`{a} di Roma, Piazzale Aldo Moro 5, I-00185 Roma, Italy}

\begin{abstract}
The quest for precision in parameter estimation is a fundamental task in different scientific areas. The relevance of this problem thus provided the motivation to develop methods for the application of quantum resources to estimation protocols. Within this context, Bayesian estimation offers a complete framework for optimal quantum metrology techniques, such as adaptive protocols. However, the use of the Bayesian approach requires extensive computational resources, especially in the multiparameter estimations that represent the typical operational scenario for quantum sensors. Hence, the requirement to characterize protocols implementing Bayesian estimations can become a significant challenge. This work focuses on the crucial task of robustly benchmarking the performances of these protocols in both single and multiple-parameter scenarios. By comparing different figures of merits, evidence is provided in favor of using the median of the quadratic error in the estimations in order to mitigate spurious effects due to the numerical discretization of the parameter space, the presence of limited data, and numerical instabilities. These results, providing a robust and reliable characterization of Bayesian protocols, find natural applications to practical problems within the quantum estimation framework.
\end{abstract}

\maketitle
\section{Introduction}
Parameter estimation is a fundamental requirement of most scientific studies. To this end, a fundamental goal consists of developing method to extract accurate information about the unknown parameters of interest from observed data. Usually, in order to achieve sufficiently high precision, measurements are repeated several times, allowing the gathering of statistical information about the studied parameters which are then treated as random variables. Measurement outcomes are indeed used to reconstruct an estimator that correctly addresses the values of the investigated parameters. The two most widely adopted and efficient estimators belong to different perspectives: the so-called frequentist approach and the Bayesian one \cite{Li_2018}. The first gives a description of the estimate as a function of the experimental outcomes and often uses the maximum likelihood estimator \cite{lane1993maximum}. The latter relies on describing the process in view of the current knowledge on the parameters, i.e., probability distributions, using both the information gained with the measurement and the prior information on the parameter values to reconstruct, through Bayes's theorem, the posterior probability distribution \cite{helstrom1969quantum}. Such a distribution is then used to give both the estimate of the parameters under investigation and of its precision \cite{kramer1988bayesian}.

Independently from the adopted approach, the interest of the metrology field is to identify which is the ultimate achievable sensitivity, intrinsically limited by the nature of the investigated process and the underlying statistical model, and which are the estimation strategies allowing the saturation of such bounds \cite{liu2022optimal}.
This brought to the investigation of different kinds of probes allowing the optimization of the estimation task. In particular, the use of quantum probes revealed to promise the capability of achieving improved estimation precision with respect to classical ones, thus disclosing applications in different tasks from biological sensing \cite{taylor2016quantum} to gravitational wave detection \cite{abbott2016observation}. For this reason, quite recently the emergent field of quantum metrology has redefined the boundaries of sensitivity, reachable when exploiting quantum resources \cite{PhysRevLett.96.010401, Giovannetti, demkowicz2014using, avsreview2020, barbieri2022optical}.

However, often the amount of available quantum resources is limited. Therefore, to achieve optimal estimation performances is necessary to properly engineer either the probe preparation or the measurement settings, or both, often based on the information acquired during the measurement process through adaptive estimation strategies \cite{PhysRevLett.104.093601,sekatski2017quantum,demkowicz2017adaptive,wiseman1995adaptive,wiseman2009quantum,higgins2007entanglement}.  This optimization process can be done through different online and offline \cite{hentschel2011efficient,lovett2013differential,cimini2019adaptive,piccoloLume,rambhatla2020adaptive} protocols. One of the most implemented ones is developed within the Bayesian inference framework, and is based on updating the knowledge of the parameter value depending on the registered measurement outcome and on the setting of some control parameter \cite{valeri2023experimental,belliardo2022optimizing,Valeri2020,cimini2023deep}. The advantage becomes particularly significant in the multiparameter framework where the interest relies on the simultaneous estimation of several parameters \cite{lee2022quantum,gebhart2021bayesian,datta2021quantum,albarelli2020perspective}. In this regime, adaptive strategies take into account the covariance structure of the parameters, optimizing the allocation of resources to estimate the parameters with the greatest precision, outperforming non-adaptive methods.

A central challenge in multiparameter estimation based on quantum strategies is the identification of an appropriate figure of merit that quantifies estimation precision and the exploration of the associated precision bounds. In this work, we analyze which are the correct figures of merit to address in the limited data regime \cite{rubio2019limited,rubio2020bayesian,meyer2023quantum} considering the Bayesian framework where both the estimation and the optimization parameters are retrieved with a discrete computation. Therefore, the appropriate figure of merit used to benchmark the quality of the estimate should also consider all the numerical caveats related to the estimation algorithm. 


Here, we examine the different figures of merit for benchmarking estimation processes in single and multiparameter quantum scenarios. Given the Bayesian framework, we consider the aforementioned caveats in the numerical calculations and the issue of the dimension of the estimation problems, within different measurement strategies. We develop a general procedure to tackle the performance benchmarking of different protocols, focusing on the median as a metric able to mitigate the presence of outliers that are intrinsic to the Bayesian estimation and to compare general protocols. A relation is demonstrated between the achieved performances and both the number of parameters and the discretization necessary to retrieve the estimate. In this way, our results provide an effective working procedure that can be adapted to different systems and scenarios to select the best-performing estimation algorithms in quantum metrology problems.

\section{Figures of merit in estimation protocols }

In the quest for the identification of the most appropriate probe and measurement strategy, a crucial analysis is to compare the achieved results with the fundamental precision bounds of the investigated process. Considering an unknown parameter $\phi$ or a vector of parameters $\vec{\phi}=(\phi_1,\phi_2,...,\phi_p)$ in a multiparameter case, a central role is played by the Cramér-Rao bound (CRB) \cite{ParisM}. Such a bound sets a lower fundamental limit on the variance (the covariance matrix $\Sigma(\vec{\phi}) $ in the multiparameter case) of unbiased estimators:
\begin{equation}
    \Delta^2\phi \ge \frac{1}{N F(\phi)}; \qquad \Sigma(\vec{\phi}) \ge \frac{\mathbb{F}^{-1}(\vec{\phi})}{N},
\end{equation}
where $N$ represents the number of employed independent probes, $F(\phi)$ and $\mathbb{F}(\vec{\phi})$ are the Fisher Information and the Fisher Information matrix respectively \cite{liu2020quantum,goldberg2021intrinsic,barbieri2023fisher}. 
Although the CRB serves as a fundamental benchmark of the achievable ultimate precision, other relevant bounds proper of the Bayesian framework have been introduced, such as the Van Trees and the Ziv-Zakai bounds \cite{trees2007bayesian, tsang2012ziv, d2022experimental}.

In the limit of a large number of probes, i.e., $N\gg1$, both the maximum likelihood estimator and the Bayesian one converge to the CRB. Indeed they are efficient estimators \cite{avsreview2020}, and therefore in this regime, the sensitivity reached by the frequentist and Bayesian methods asymptotically agrees.  However, in real experimental conditions, it is usually desired to exploit the minimum number of probes to show such a convergence. This is possible by optimizing the measurement and the probe preparation in order to extract the highest possible amount of information from each probe. This allows a faster convergence to the bound and the possibility of saturating it already in the limited resource regime.

In the quantum scenario, implementing such optimization protocol, however, is not a trivial task in particular for multiparameter problems. Indeed, the solution of the optimization problem requires in general expansive computational efforts since it involves the computation of complex multidimensional integrals. To speed up such computation an efficient solution is represented by the Sequential Monte Carlo (SMC) \cite{granade2012robust} algorithm which has indeed been already employed to solve different problems from Hamiltonian learning \cite{granade2012robust,wiebe2014hamiltonian}, to state tomography \cite{huszar2012adaptive} and parameter estimation \cite{valeri2023experimental,Valeri2020,wiebe2016efficient} often using an adaptive scheme. 

The precision of an estimator in quantum metrology is usually assessed in terms of its variance. In the Bayesian scenario, this figure of merit corresponds to the variance of the posterior distribution $P(\phi|x)$ updated after the measurement outcome $x$:
\begin{equation}
    \sigma^2 = \int d\phi P(\phi|x) [\tilde{\phi}(x)-\phi]^2,
\end{equation}
where $ \tilde{\phi}(x)$ represents the estimate of the parameter, obtained as follows:
\begin{equation}
    \tilde{\phi}(x) = \int d\phi P(\phi|x) \phi.
\end{equation}
However, the variance alone may not fully capture the performance of the estimator since it is related only to the spread of the estimator probability distribution, and it does not take into account potential systematic errors or biases. To face this issue, in practical applications, another figure of merit that is often taken into consideration, allowing a more comprehensive evaluation of the achieved estimation performances, is the quadratic loss. The latter is defined as the quadratic difference between the estimated parameter $\tilde{\phi}$ and its true values $\phi^{true}$ ($\phi_i$ and its true values $\tilde{\phi}_i$ for the multiparameter case):
\begin{equation}
 Q(\phi)=  (\tilde{\phi} - \phi^{true})^{2}; \qquad   Q(\vec{\phi})= \sum_{i=1}^p (\tilde{\phi}_{i} - \phi^{true}_{i})^{2}.
\end{equation}
The quadratic loss naturally embeds both the variance and the bias in the estimator, often resulting in a more robust quantity to evaluate the estimation performances. While in actual experiments the true values of the parameters are unknown and thus the quadratic loss is inaccessible, this quantity is a precious tool to pre-calibrate the performances of estimation strategies, for instance through numerical simulations that take into account a detailed model of the apparatus.


\section{Benchmarking quantum metrology strategies}
\begin{figure}[ht!]
\centering
\includegraphics[width=0.99\columnwidth]{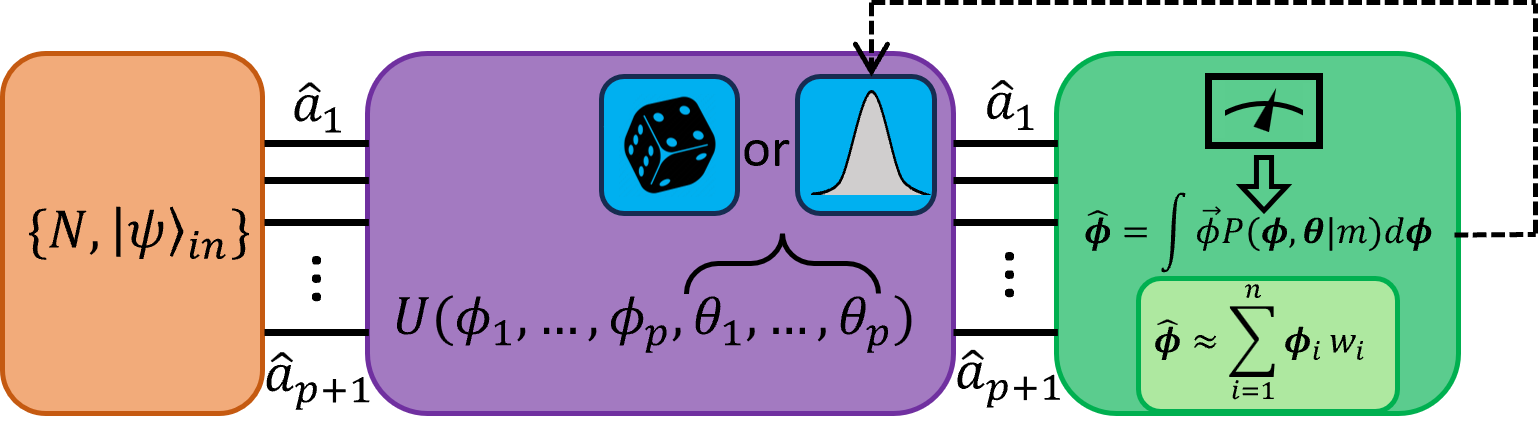}
\caption{Schematic of the studied estimation problems. $N$ probe states $|\psi\rangle_{in}$ are prepared and evolves following the unitary $U$. For the single parameter case, the probe is a single-particle state while for the two- and three-parameter problems is a two-particle state. The evolution depends on the $p$ parameters of interest $\{\phi_1,...,\phi_p\}$ and on a set of control parameters $\{\theta_1,...,\theta_p\}$ that have the same effect of shifting the measurement point. These control parameters can either be set randomly or by employing a feedback-based adaptive strategy. The measurement outcomes $\{m\}$ allow us to reconstruct the Bayesian estimator, using the particle filtering approximation that employs $n$ particles.}
\label{fig:s}
\end{figure}

We now focus on practical case studies of Bayesian quantum metrology estimations, within which we study the performances of protocols as a function of the relevant parameters of the processes by using different figures of merit.
To this goal, we consider a test-bed example starting from the estimation of the rotation angle $\phi$ of a single-qubit state. Then, we enter in the multiparameter regime studying two- and three-parameter estimation problems with two-particle probe states. 
The general estimation process considered here is schematized in Fig. \ref{fig:s}. 

\subsection{Single parameter case}

We start considering the estimate of $\phi$ encoded in the single-qubit state $|\psi(\phi)\rangle = 1/\sqrt{2}\big(|0\rangle\pm e^{i\phi}|1\rangle\big)$, with $\phi\in[0,2\pi)$. This coincides also, in the optical case, with the output single-photon state of a two-arm interferometer. We study the performances achieved with numerical simulations when adopting a Bayesian estimator updated after each measurement outcome. Quantum parameter estimation protocols featuring non-monotonic likelihood probabilities, such as most phase estimation problems, have an ambiguity issue. Indeed, a fixed probability can be generated by more than one phase even if the parameters are inside the same periodicity interval. In these cases, it is necessary to implement strategies that can distinguish between these ambiguities, and this can be achieved by changing the measurement settings during the estimation process. For this purpose, and also for faster convergence of the estimation to the precision bound, adaptive protocols can be used \cite{wiseman1995adaptive,berry2000optimal}. Different adaptive algorithms are characterized, in general, by different performances. 

\begin{figure}[ht!]
\centering
\includegraphics[width=0.99\columnwidth]{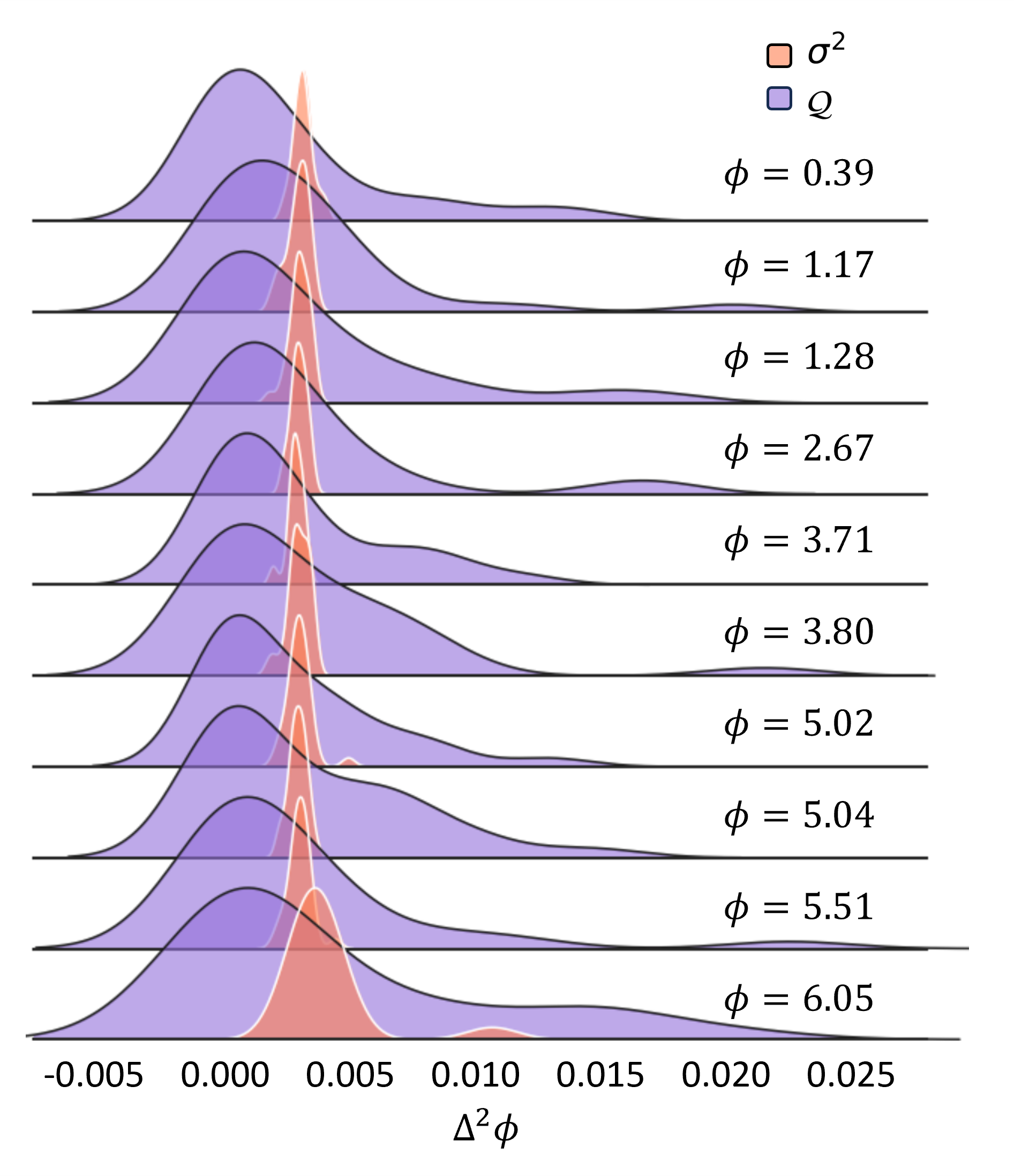}
\caption{Normalized probability distribution of the performances of the estimation algorithm over different repetitions. The performances are reported in terms of quadratic loss values (purple) and variances (orange) obtained over the $30$ different repetitions for each phase value $\phi$, after sending $N=300$ probes. The distributions are obtained with the kernel density estimation \cite{turlach1993bandwidth} which gives the estimated probability density function smoothed on the histograms of the data. The performances are reported for $10$ selected values among the $100$ tested ones in order to cover the whole periodicity interval.}
\label{fig:2}
\end{figure}

In order to make the results of this work independent of the specific optimization protocol chosen, we focus only on the easiest and computationally efficient protocol that consists of shifting the measurement settings randomly after the measurement of each probe. More specifically, to disambiguate the phase values in the whole periodicity interval, it is necessary to adopt a procedure that changes the measurement settings during the estimation experiment. The easiest and computationally efficient one is to shift the measurement settings randomly after the measurement of each probe. This random control-setting strategy permits to saturate the CRB in the single parameter estimation. The reported performances are obtained considering $M = 100$ different rotation angles $\phi$ chosen randomly in $[0,2\pi)$ and repeating the estimate of each phase $r = 30$ times to study its robustness. We study the performances as a function of the number of probes $N$, and we compare them with the CRB that provides a lower bound on the achievable estimation precision. The performances are computed both in terms of variance and quadratic loss and the distributions, obtained from the respective histogram over the $30$ repetitions, for $10$ of the $100$ investigated phases are reported in Fig.\ref{fig:2}. As it appears from the comparison of the distributions of different repetitions for each rotation angle, the influence of the outliers on the mean value of the quadratic loss is greater. This is observed from the tails of the distributions, while those of the variance are more shrunk around a central value. 

\begin{figure}[ht!]
\centering
\includegraphics[width=0.99\columnwidth]{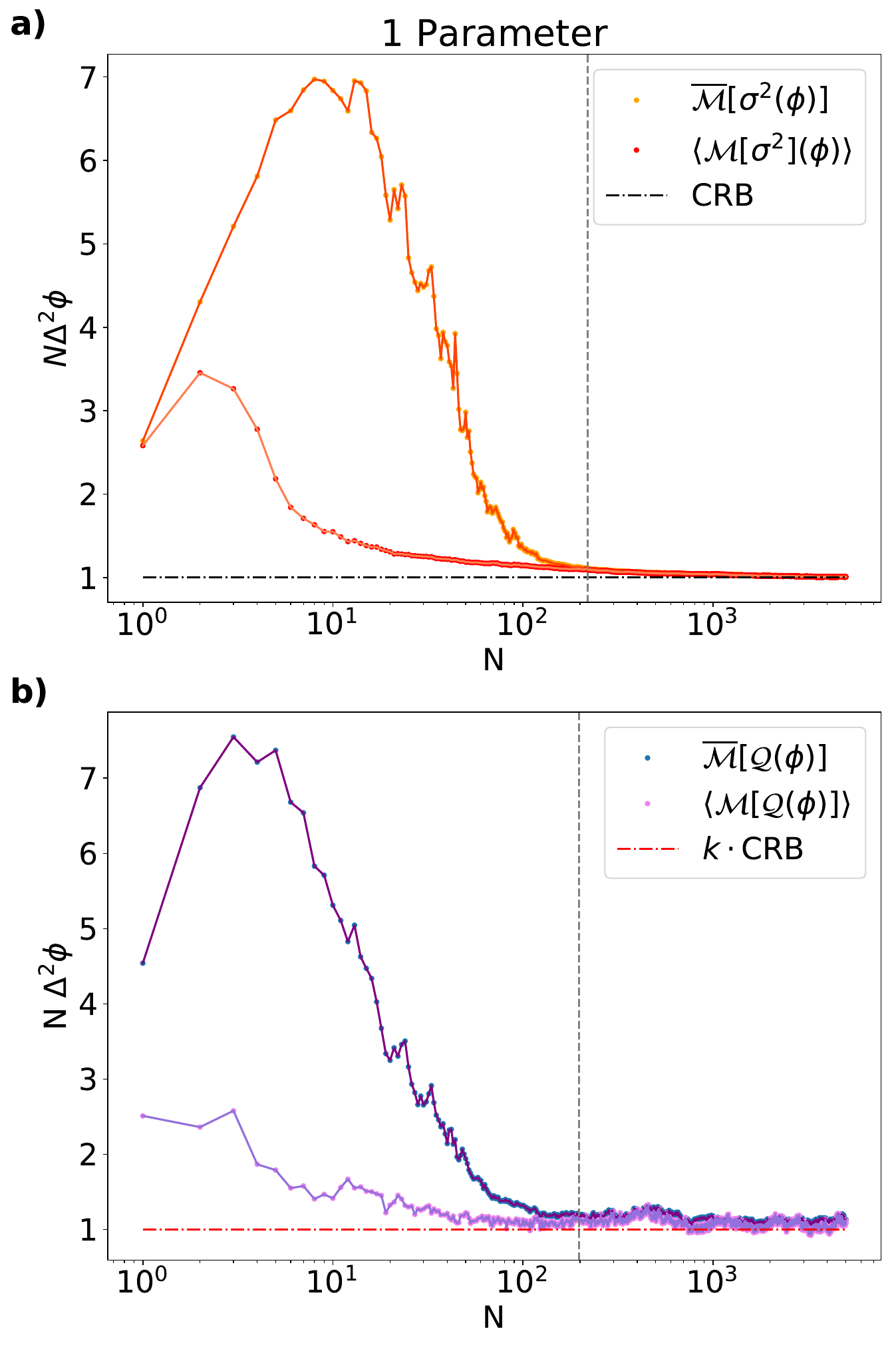}
\caption{\textbf{a)} Comparison of estimation performances in terms of variance with the CRB considering the mean (orange points) over the different $100$ angles inspected or the median (red points) on the overall sample. \textbf{b)} Comparison of estimation performances in terms of quadratic loss with the CRB considering the mean (blue points) over the different $100$ angles inspected or the median (violet points) on the overall sample. The gray dashed line represents the point from where the ratio among the two quantities $\frac{\overline{\mathcal{M}}}{\langle{\mathcal{M}}\rangle}$ becomes lower than the $1\%$.}
\label{fig:6}
\end{figure}

The outliers above can impact the benchmarking procedure of a specific strategy. For this reason, it can be useful to replace the analysis of the mean with the median of the set of different estimation runs. The different behavior of the distribution is due to the fact that for $N$ sufficiently large, the posterior becomes a Gaussian distribution centered around the true value of the parameter and with variance given by the inverse of the Fisher information. It follows that the variance is Gaussian distributed. In this case, the mean and the median coincide, and the bound of the two figures of merit remains the CRB. This is not the case for the quadratic loss, which is the square of a variable described by a Gaussian distribution, resulting in a positively skewed distribution. In this case, the mean is greater than the median, and the proportionality factor among them can be estimated numerically, and in the single-parameter case it results equal to $k\simeq 0.4549$ \cite{belliardo2022optimizing}. This factor must be taken into consideration when comparing the performances in terms of medians with the ultimate precision bound. 
Moreover, we underline that when comparing the overall estimation performances to the precision bound, after having considered either the mean or the median over the different repetitions of the estimate (that we indicate respectively with $M[\cdot]$ and $\mathcal{M}[\cdot]$), it is necessary to consider the average performances among all the different parameter values. Also in this case, there are some differences in the considered figure of merit but only in the range of very low probes as reported in Fig.\ref{fig:6}. Hereafter, we indicate the mean over different angles as $\overline{\phi}$ while the median as $\langle\phi\rangle$.

\begin{figure}[ht!]
\centering
\includegraphics[width=0.99\columnwidth]{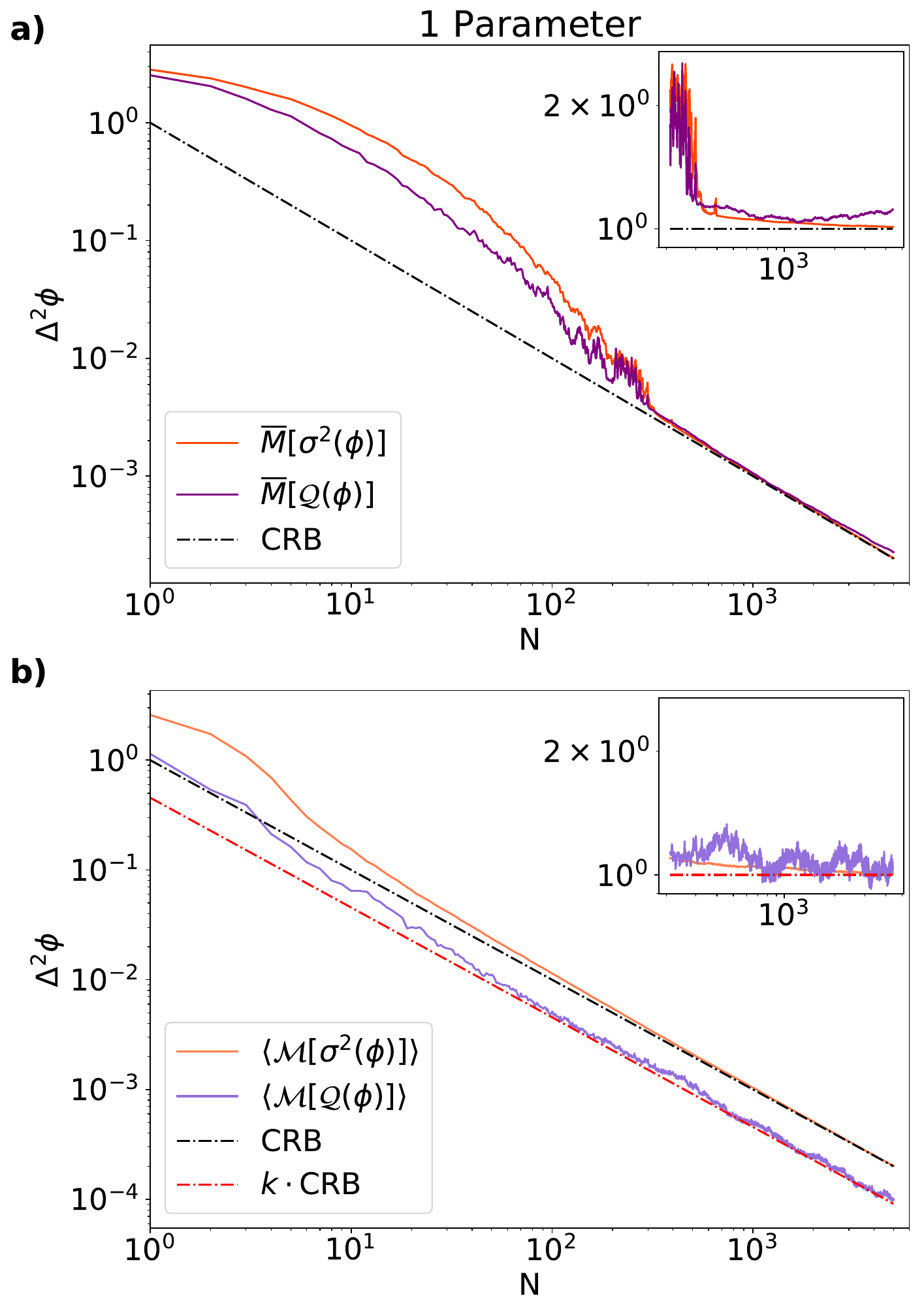}
\caption{Performance of single parameter estimation achieved with the random adaptive strategy. \textbf{a)} The performances are reported averaging the obtained variances (dark orange) and quadratic losses (purple) over the $100$ different parameter values investigated and the $30$ repetitions for each one. The mean of the quadratic loss and variance is reported as a function of the number of probes $N$ employed and it is compared with the CRB (black dashed line). \textbf{b)} The performances are reported by computing the median over the different parameter values and repetitions of the variances (coral) and of the quadratic losses (violet). While the median of the variances is still bounded by the CRB, the median of the quadratic loss is bounded by a rescaled bound $k\cdot$CRB (red dashed line). The two insets show the same values weighted for the relative number of probes and for the factor $k$ starting from $N=200$.}
\label{fig:1}
\end{figure}

Due to the shape of the quadratic loss distribution, it becomes more sensitive to outliers. The conventional reliance on the mean estimate across different repetitions of the estimate and parameter values can be problematic as it may not converge to the CRB, largely because of the impact of these outliers. This may thus include a bias when benchmarking or comparing different techniques. This is shown in panel a) of Fig.\ref{fig:1} where it is reported the mean of the achieved variance and quadratic loss computed over the same set of data. The median, on the other hand, offers distinct advantages, especially in scenarios where resources are limited. By adopting the median and considering the associated bound, as reported in panel b) of Fig.\ref{fig:1} we observe a clear convergence to the bound, also for the quadratic loss, with fewer probes compared to using the mean estimate. In such a way, using the median provides an approach to identify the convergence to the ultimate bound of a given estimation algorithm by reducing the impact of outliers.

\begin{figure*}[ht!]
\centering
\includegraphics[width=0.99\textwidth]{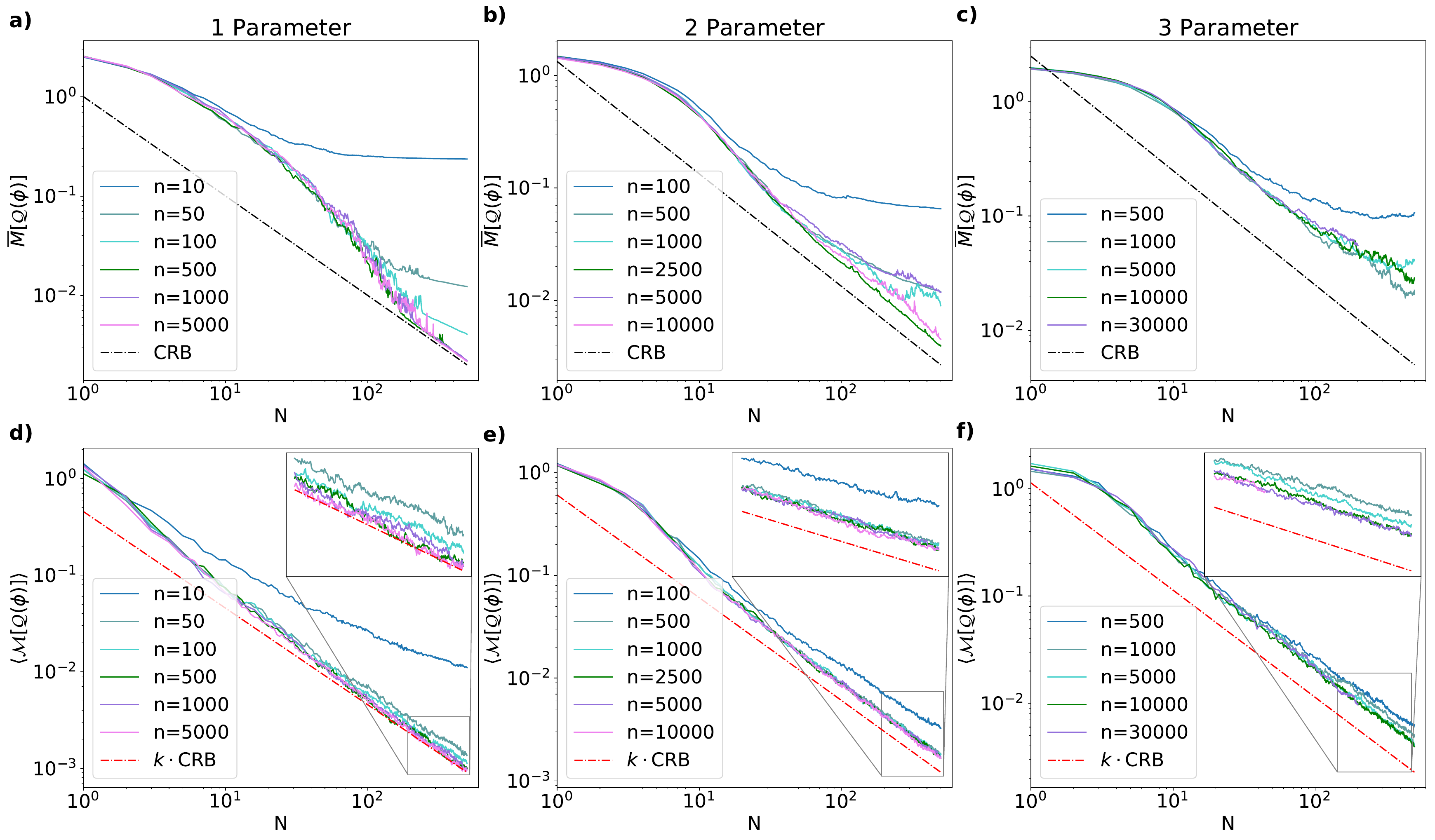}
\caption{Comparison of the performances of the single- and multi-parameter estimation algorithms in terms of quadratic loss for different discretizations achieved adopting for all the random adaptive strategy. In each plot, the different curves represent the achieved quadratic losses changing the number of overall particles $n$ in the estimation algorithm. The plots in the first row represent the mean of the quadratic loss over the $100$ different angles and the $30$ repetitions for each with the CRB (black dashed line), while the ones in the second row report the medians and its rescaled CRB (red dashed line).}
\label{fig:3}
\end{figure*}

As previously discussed, our approach to computing the necessary integrals for Bayesian estimation involved discretizing the parameter space into $n$ possible values covering the entire periodicity interval. This discretization was conducted in accordance with SMC, also known as the particle filtering method \cite{granade2012robust}, used in recent multiparameter estimation experiments \cite{Valeri2020,valeri2023experimental,belliardo2022optimizing}. The objective was to reconstruct the posterior probability distribution of the parameter and then use it to estimate its value. It is important to recognize that the convergence to the ultimate bound of precision is closely tied to the choice of the number of particles $n$ used in this process.  The number of particles employed in the estimation process impacts how well the parameter space is sampled and how effectively the estimation algorithm converges. A higher number of particles can improve precision, but it also increases computational requirements. Therefore, the selection of the appropriate number of particles is not straightforward but rather involves a trade-off between reliability and computational cost that in turn increases with the number of parameters. The convergence to the bound will therefore be related also to such choice. Indeed, the number of particles affects the ultimate achievable precision that in a first approximation, and in the absence of resampling, can be expressed as $\big(\frac{2\pi}{n}\big)^p$, where $p$ indicates the number of parameters being estimated. The performances expressed as both the mean and the median of the obtained quadratic losses are reported for different discretization choices for single- and multi-parameter problems in Fig.\ref{fig:3}. The obtained performances overcome the simple discretization limitation since all the simulations are done employing resampling. The influence of finite discretization results in the attainment of a plateau value, corresponding to the limited sensitivity achievable with such discretization, as it appears in the plots of the first row of Fig.\ref{fig:3}. In these plots the average achieved quadratic losses are reported as a function of the number of injected probe states. The effect of discretization of the integration space is attenuated when considering instead the median of the achieved results, providing also in this case a more robust and reliable measure of the general performances achieved by the estimation algorithm, filtering out fluctuations arising from the finite discretization. 

\subsection{Multiparameter case}

We now consider a second scenario, involving the estimate of two or three phases embedded in a multiarm interferometer injected with two-photon states following \cite{Valeri2020,valeri2023experimental}.  Starting from the scheme reported in Fig.\ref{fig:1}, the applied transformation in the considered single and multiparameter scenarios is $ U(\vec{\phi},\vec{\theta}) = e^{i \sum_{k=1}^p \hat{n}_k(\phi_k+\theta_k)}$, with $\hat{n}_k$ the particle number operator of the $k$ mode. As for the single parameter case, we first study the strategy where the controls are varied randomly during the estimation process. As observed in Fig.\ref{fig:3}, we find that while the bound is saturated for the single parameter problem, for the analyzed system, this is not the case in the multiparameter framework. The reason why it remains a bias from the achieved quadratic loss and the bound is related to the choice of the adaptive protocol. 
In the multiparameter scenario, the shape of the likelihood function, and as a consequence the Fisher information landscape, is more complex, and not flat over the parameter space. Therefore in order to disambiguate among different parameter values in a limited-data regime, a more efficient adaptive strategy involving a feedback loop between the measurement outcomes and the experimental settings is required. 

\begin{figure}[ht!]
\centering
\includegraphics[width=0.99\columnwidth]{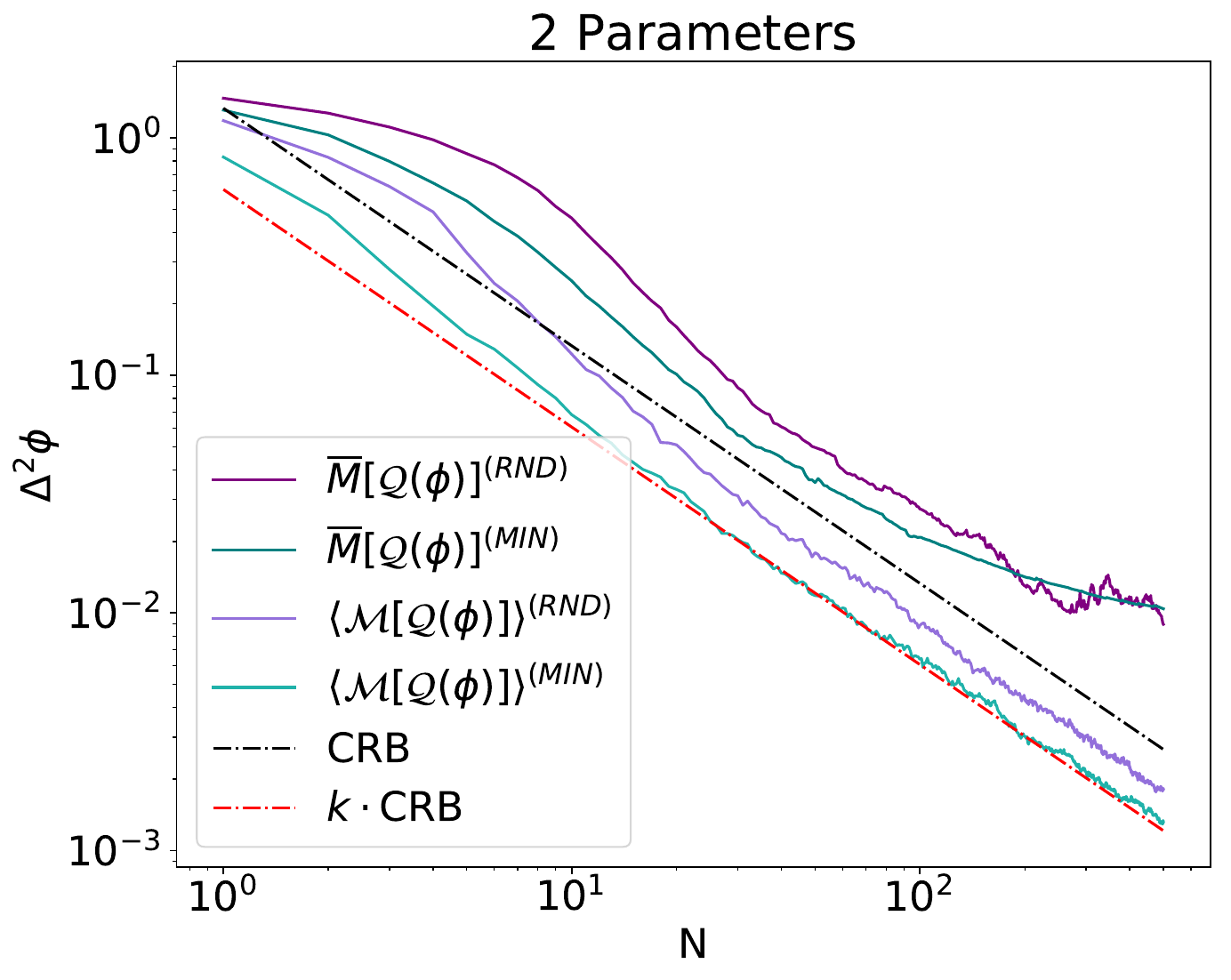}
\caption{Comparison of the random adaptive strategy and the one setting the control parameters with a feedback-loop depending on the measurement results for the two-parameter scenario. The estimation performances are reported in terms of quadratic loss as a function of the number of probes $N$ with a number of particles set to $n=1000$. The results obtained with the random strategy are reported in terms of mean (purple line) and median (violet), as well as the ones obtained with the minimization algorithm for which the mean (cyan) and the median (light blue) are also computed.}
\label{fig:4}
\end{figure}

The realm of possible adaptive techniques is varied. Among the others, it includes machine learning-based algorithms \cite{fiderer2021neural,cimini2023deep}, variational techniques \cite{cimini2023variational,meyer2021variational}, Bayesian updates based on minimizing the variance of the posterior \cite{valeri2023experimental,Valeri2020,belliardo2022optimizing}, particle swarm optimization \cite{hentschel2011efficient}, genetic algorithms \cite{rambhatla2020adaptive}, or differential evolution \cite{lovett2013differential,palittapongarnpim2019robustness}. 
The choice of one of these efficient adaptive protocols, involving a feedback loop, becomes crucial to achieve the ultimate sensitivity in multiparameter scenarios. This is often necessary to disambiguate multiple parameters in a limited-data regime with increased complexity of the likelihood function. This is evident in Fig.\ref{fig:4} where, already for the two-parameter problem, it is necessary to adopt a minimization algorithm with a feedback loop to select the measurement settings. This is required to converge towards the ultimate precision bound. Indeed in this regime, the parameters under estimation often exhibit correlations, where changes in one parameter can significantly impact the estimation of others resulting in interdependent parameter uncertainties. Adaptive strategies allow for dynamic adjustments of measurements, taking into account the evolving estimates of all parameters, thereby optimizing resource allocation and minimizing the trade-offs among the uncertainties associated with each parameter \cite{albarelli2020perspective}.

In this scenario, utilizing the median as a metric of merit, as opposed to the mean, offers distinct advantages in benchmarking the performances of an estimation strategy also for a reduced discretization. Specifically, when dealing with a system where the Fisher information exhibits divergences, it becomes increasingly challenging to achieve high-precision estimations for those points in close proximity to the divergent regions, especially when the number of particles is not sufficiently high. As a result, the effectiveness of the minimization algorithm becomes strongly contingent on the specific parameter values being considered. Consequently, if the estimation algorithm does not perform optimally for only a small subset of the $100$ phases under investigation, it significantly impacts the overall performance when assessed in terms of the mean. This issue can be mitigated by examining the median, which helps reduce the influence of these outliers, allowing to enhance the robustness of the benchmarking process against the constraints imposed by finite discretization, which are linked only to numerical limitations. Therefore, also in this case, the median provides a more accurate representation of the estimation quality itself.

Note that the goal here is to test estimation techniques in a robust way, mitigating the presence of outliers and instability effects due to numerical discretization. This is a different problem with respect to the case where one considers the performances of a single measurement. Here, the outliers have a non-zero probability to occur, thus having an impact in the single estimation process.

\begin{figure}[htb!]
\centering
\includegraphics[width=0.99\columnwidth]{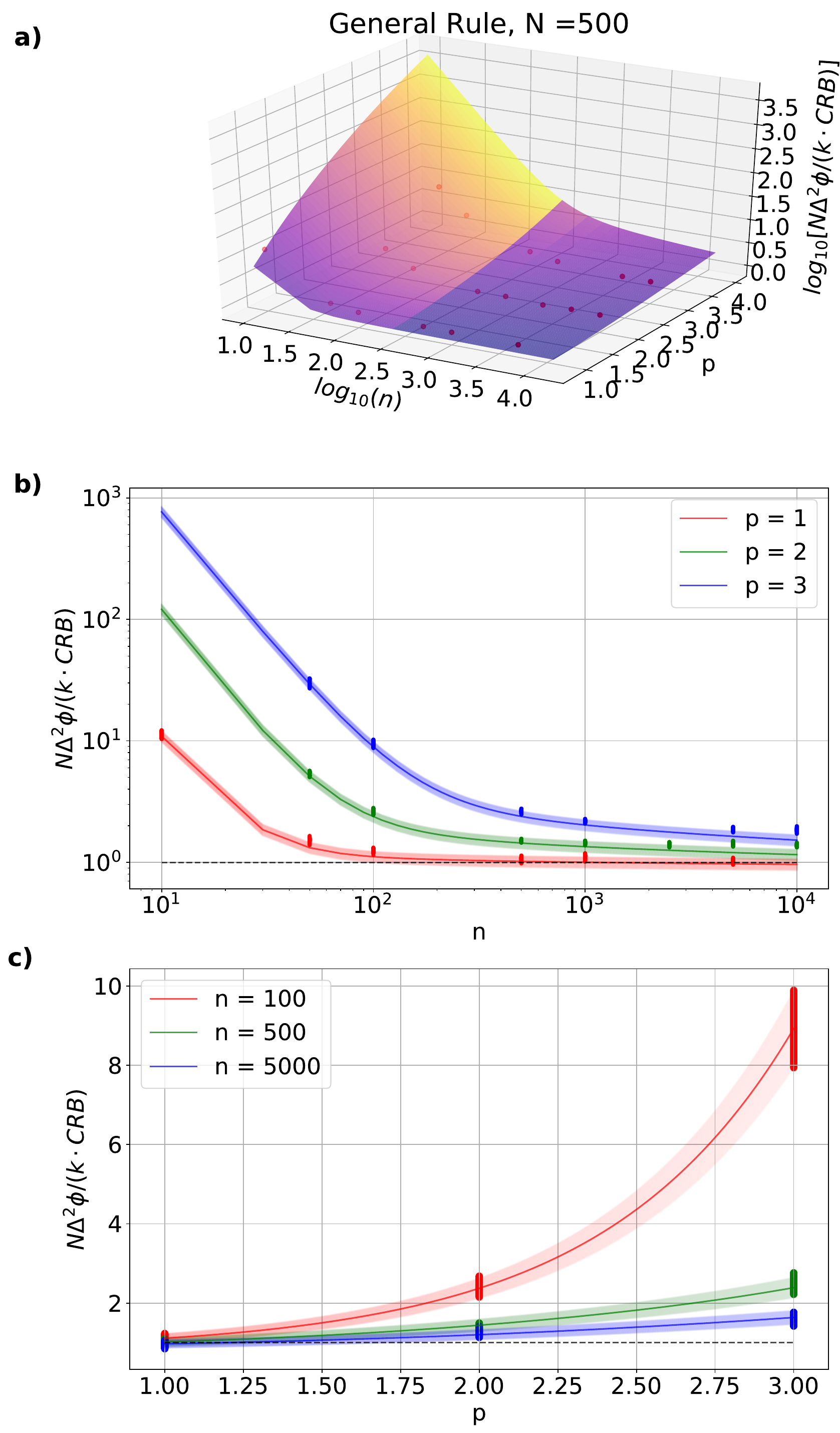}
\caption{General rule describing the median of the achieved quadratic loss as a function of the number of particles $n$ and of the number of parameters of interest $p$. The results are reported rescaling the quadratic loss with the number of probes employed and the relative bound in order to show the convergence to one independently from the specific problem considered. \textbf{a)} Fit on the obtained numerical results (red dots) when fixing the number of probes to $N=500$. \textbf{b)} Trend of the median quadratic loss versus the number of particles $n$ when fixing the number of parameters to $p=1,2,3$. \textbf{c)} Trend of the median quadratic loss versus the number of parameters $p$ when fixing the number of particles to $n=100,500,5000$. Plots \textbf{b)} and \textbf{c)} show the results when varying the number of probes from $N=400$ to $N=500$. The dashed line represents the value of 1, corresponding to the saturation of the CRB.}
\label{fig:5}
\end{figure}


\section{Practical guidelines}
Driven by the above results, we discuss below heuristic guidelines for general estimation problems. It becomes beneficial for practical implementations to have a working procedure that allows the identification of the best performing estimation protocol. However, to perform a fair comparison among the adopted approaches it becomes important to avoid alteration due to outliers that are intrinsic to the shot-by-shot Bayesian procedure. Therefore, defining a rule for the considered system that provides a systematic framework for achieving the minimum uncertainty while taking into account constraints such as the number of probes and the level of discretization become crucial in this kind of protocol. This can help in the assessment of resources for specific precision goals, optimizing these parameters to achieve the best trade-off between achievable precision and available resources.

As the number of parameters increases, it becomes important also to deal with the presence of possible correlations among their uncertainties, meaning it becomes more challenging to achieve precise estimates for all the parameters simultaneously. The challenge is compounded by the fact that the number of particles required to show a convergence to the ultimate bound also grows with the number of parameters. The median unique feature of smoothing these trends, reducing the effects of specific phase values, enables the derivation of a generic model describing how the precision is related to the number of parameters and the number of particles employed in the discretization process. The trends retrieved are shown in Fig.\ref{fig:5} where the relation among the distance from the achieved uncertainty and the relative bound is reported for different numbers of particles $n$ and for problems with a different number of investigated parameters $p$. Notably, these trends are verified independently from the number of probes employed, provided that their number is sufficiently large to grant the saturation to the bound in the adaptive configuration.

To obtain a rule governing the performances scaling of the considered system, we have performed a multidimensional fit of the median of the obtained quadratic loss over the last $100$ probes i.e. $400\le N\le 500$. The results are then adjusted for the relative bound, which varies depending on the number of parameters being estimated. The considered inputs for the fitting process are the number of investigated parameters and the number of particles employed in the Bayesian estimate.    
The function retrieved with the heuristic process is:
\begin{equation}
    f(n,p) = A (1 + \sqrt{p})^{2} p \, n^{(- B + C \, p)} + D \sqrt{p} + E \frac{p^{F}}{n^{G}},
\label{eq:fittyno}
\end{equation}
with
\begin{equation}
\begin{split}
    &A \simeq 160, \quad B \simeq 5.4, \quad C \simeq 0.8, \quad D \simeq 0.002,\\ &E \simeq 0.11, \quad F \simeq 3.5, \quad G \simeq 1.3.
\end{split}
\label{eq:paramelli}
\end{equation}
This function has been obtained by making considerations on the exponential scaling of the space of the discretization grid and its relation with the number of parameters i.e. the factor $n^{-b+c \, p}$ linked with the scaling of the quantum fisher information with the number of parameters \cite{liu2020quantum} giving the factor $a (1+\sqrt{p})^2p$. Other terms are instead needed in order to take into account the specific behavior of the likelihood of the investigated system and the impossibility of disambiguating all the phase values with the random strategy.


We note that this work focused on phase estimation protocols. Nevertheless, we expect that the approach can be adapted to general Bayesian estimation problems. The same issues we studied will affect general Bayesian protocols and the expression in Eq. \eqref{eq:fittyno} can likely describe the performances as a function of the number of particles, parameters and probes, with suitable parameters values, that can be different from those in Eq. \eqref{eq:paramelli}.

\section{Discussions}

The aim of this work was to analyze how to robustly benchmark Bayesian estimation protocols, addressing both single and multiparameter scenarios. In the Bayesian framework, an essential role is played by numerical calculations that approximate the belief propagation and allow for optimized protocols such as adaptive ones. Moreover, any  practical estimation process has to face finite statistics, in which only a limited amount of resources are used. In this context, a crucial task for quantum metrology is to benchmark the optimality of different estimation protocols. A primary question is to find the right figure of merit able to assess and compare the considered approaches. 

In order to highlight the most effective estimation protocol we have performed a comparative analysis, demonstrating the benefit of using the median of quadratic error as a metric to assess the estimation performances, mitigating effects from numerical discretization, limited data, and outliers inherent in Bayesian estimation. This choice becomes pivotal for benchmarking multiparameter estimation protocols where complexities arise from interdependencies between parameters. Quantifying the relation between uncertainty and number of parameters and particles revealed a heuristic model capturing performance trends. This model aids in resource optimization, enabling trade-offs between achievable precision and computational constraints.

This approach offers a pathway for optimizing resource allocation and achieving optimal precision in estimation processes. In this way, our analysis represents a tool with direct application to disparate metrology problems and platforms: from gravitational wave detection\cite{abbott2016observation}, biological sensing \cite{taylor2016quantum}, sensing with Gaussian states \cite{morelli2021bayesian}.

\section*{Acknowledgments}
This work is supported by the PNRR MUR project PE0000023-NQSTI (Spoke 4 and Spoke 7). N. S. would like to acknowledge funding from Sapienza Università di Roma via Bando Seed PNR 2021, Project AQUSENSING (Advanced Calibration and Control of Quantum Sensors via Machine Learning).

\bibliography{biblio.bib}

\end{document}